# Measuring the Overall Complexity of Graphical and Textual IEC 61131-3 Control Software

Juliane Fischer[1], Birgit Vogel-Heuser[1], Heiko Schneider[1], Nikolai Langer[2], Markus Felger[3], Matthias Bengel[3]

*Abstract*—Software implements a significant proportion of functionality in factory automation. Thus, efficient development and the reuse of software parts, so-called units, enhance competitiveness. Thereby, complex control software units are more difficult to understand, leading to increased development, testing and maintenance costs. However, measuring complexity is challenging due to many different, subjective views on the topic. This paper compares different complexity definitions from literature and considers with a qualitative questionnaire study the complexity perception of domain experts, who confirm the importance of objective measures to compare complexity. The paper proposes a set of metrics that measure various classes of software complexity to identify the most complex software units as a prerequisite for refactoring. The metrics include complexity caused by size, data structure, control flow, information flow and lexical structure. Unlike most literature approaches, the metrics are compliant with graphical and textual languages from the IEC 61131-3 standard. Further, a concept for interpreting the metric results is presented. A comprehensive evaluation with industrial software from two German plant manufacturers validates the metrics' suitability to measure complexity.

## I. COMPLEXITY MEASUREMENT WITH METRICS

AUTOMATED Production Systems (aPS) form a special class of mechatronic systems, namely automated machines and plants, whose development involves mechanics, electrics/ electronics and software, all closely interwoven [1]. Thereby, control software implements an increasing proportion of aPS functionality [2], [3] and is programmed with the graphical and textual languages defined in IEC 61131-3, usually executed on Programmable Logic Controllers (PLCs). Apart from pure functional tasks like actuator control, PLC software also includes extra-functional parts, e.g., different operation modes and fault handling. Thus, it is highly influenced by automation hardware, customer requirements and even country-specific laws, leading to high complexity in aPS control software.

To compete on the global market and meet requirements coming from Industry 4.0, e.g., lot-size one, software needs to be reused to shorten the development time. Further, high-quality software is required and its maintenance needs to be eased by design. However, oftentimes the control software's high complexity reduces its comprehensibility, reusability and maintainability, thus leading to an increased effort during its development, testing and maintenance [4], [5], [6]. Project planning would benefit from quantifying the additional costs caused by complexity but this is not yet feasible, as multiple, different and highly subjective views on software complexity exist, ranging from the pure size of the software to its comprehensibility. Although there are standards defining software quality, e.g., ISO/IEC 25010, these different views and definitions make it difficult to measure the complexity of control software. To ease reuse and, thus, reduce costs, identifying highly complex software parts for their subsequent refactoring is essential, as their improvement is expected to generate the greatest benefit. For this purpose and to estimate the added value of the refactoring effort, objective measures for software complexity considering the different views are required. Although various software complexity metrics are available, they only focus on single views of complexity or are not applicable to textual and graphical IEC 61131-3 languages.

This paper's main contribution is a metric, applicable to graphical and textual programming languages, to compare IEC 61131-3 control software with respect to its overall complexity considering five complexity classes, which enables identifying the most complex software parts as a starting point for refactoring. Domain experts confirm the relevance of the complexity classes and the need for objective complexity measures in a qualitative study and the metric is evaluated using 50 software units from two German plant manufacturers. The

Manuscript received: February 17, 2021; Revised: May 1, 2021; Accepted: May 24, 2021.This paper was recommended for publication by Editor Jingang Yi upon evaluation of the Associate Editor and Reviewers' comments. This work was supported by *Bayerische Staatsministerium für Wirtschaft und Medien, Energie und Technologie* under Grant DIK0112.
[1]J. Fischer, B. Vogel-Heuser, and H. Schneider are with the Institute of Automation and Information Systems at Technical University of Munich, Boltzmannstr. 15, 85748 Garching, Germany (e-mail: {juliane.fischer; vogel-heuser; heiko.schneider}@tum.de).
[2]N. Langer is with Brückner Maschinenbau GmbH & Co. KG, Königsberger Str. 5-7, 83229 Siegsdorf, Germany (e-mail: nikolai.langer@brueckner.com).
[3]M. Felger and M. Bengel are from teamtechnik Maschinen und Anlagen GmbH, Planckstraße 40, 71691 Freiberg, Germany ({Markus.Felger; Matthias.Bengel}@teamtechnik.com).
Digital Object Identifier (DOI): see top of this page.

remainder of this paper is structured as follows: Sec. 2 presents derived requirements for complexity measurement, followed by an introduction to the state of the art in Sec. 3. Then, Sec. 4 summarizes insights gained from industry through a qualitative questionnaire study. Sec. 5 introduces the complexity metrics and results visualization concept, and next, Sec. 6 gives insights on the metrics' evaluation using industrial control software. The paper closes with a summary and outlook.

## II. DERIVED REQUIREMENTS FOR COMPLEXITY METRICS

The requirements targeted within this paper are derived in the following from current challenges in the aPS domain.

Since software implements a significant proportion of functionality in aPS, its reuse is essential to stay competitive. However, understanding and, thus, reusing complex software units is not a trivial task. Objective measures, e.g., software metrics, can help identify complex software units as the first step to refactoring to support reuse. Since various views on and reasons for software complexity are known, different classes of complexity should be considered in the complexity assessment. Therefore, metrics for measuring control software complexity from different views need to be derived and/or developed and combined to gain an overview of aPS software (*Requirement R1 considering different complexity views*).

The IEC 61131-3 standard defines textual and graphical programming languages, which have different strengths, e.g., Ladder Diagram (LD) is suitable for programming interlocks, while Structured Text (ST), which is similar to Pascal, is used to implement mathematical operations. Due to their different characteristics, the languages are often combined within a control software project since different languages are suitable to program different automation tasks. For complexity measuring in aPS control software, all IEC 61131-3 languages need to be considered to obtain a complete overview of the analyzed project. Thus, Requirement 2 is that complexity metrics need to be applicable to textual and graphical IEC 61131-3 languages (*R2 textual and graphical languages*).

Further, it needs to be ensured that the developed software metrics are applicable to industrial control software to address current challenges in the aPS domain and support its reuse. Only using the complexity measurement concept in industry can ensure that the metrics take the domain's boundary conditions into account. Furthermore, the scalability of the metrics regarding the size of industrial control software (up to 270 or more software units [2], some of which have up to 1500 Source Lines of Code (not counting comments)) needs to be ensured. Thus, Requirement 3 targets the complexity metrics' applicability to industrial control software (*R3 scalability*).

Automated static code analysis is an efficient solution for identifying disadvantageous parts in the early development stages [7]. Further, the static analysis of source code is a means to quantify software attributes such as quality. However, due to the size of aPS control software, a manual calculation of metrics is not feasible and its effort would not outweigh the benefits of identifying complex software parts. Consequently, the developed complexity metric's result needs to be determinable using static code analysis with as little effort as possible to make its application beneficial (*R4 automatable*).

Another important aspect regarding static code analysis and software metrics is the comprehensibility of the results [8]. Software metrics for measuring complexity are only useful if the developer can understand and interpret the obtained results and, in the best case, directly identify points for improving the software. Thus, the metric results obtained should provide information on possible recommendations for action and be understandable for the metric user (*R5 comprehensibility*).

## III. CONTROL SOFTWARE DEVELOPMENT AND APPROACHES FOR ITS COMPLEXITY MEASUREMENT

This section introduces aPS software development, including software quality and metrics. Further, related work from computer science and metrics developed for complexity measurement in IEC 61131-3 control software are presented.

### A. Software Development in the Domain of aPS

aPS are usually controlled by PLCs, which are characterized by a cyclic program execution with fixed cycle times to ensure process stability. To structure control software into reusable parts through functionality encapsulation, IEC 61131-3 defines three types of Program Organization Units (POUs), namely Functions (FCs), Function Blocks (FBs) and Programs (PRGs). Each POU consists of a declaration part to define its variables and an implementation part, which is programmed in one of the five IEC 61131-3 programming languages. Three graphical (Ladder Diagram (LD), Function Block Diagram (FBD) and Sequential Function Chart (SFC)) and two textual languages (Structured Text (ST) and Instruction List (IL), with the latter rarely being used today) are defined. Different platform suppliers offer PLCs programmable in accordance with IEC 61131-3. Others, like Siemens, use slightly different constructs, e.g., Organization Blocks (OBs) instead of PRGs, but state their compliance to IEC 61131-3 [9].

By analyzing PLC software projects, five architectural levels have been identified [2]. Each level contains software modules (an individual POU or a group of POUs), ranging from *plant modules* controlling whole production plants to *atomic basic modules* for single actuators. The software is often modularized according to aPS hardware or the functionality it implements, with a modular software structure having a positive impact on software quality and facilitating reuse and extensibility [10]. Frequently, control software is divided into standardized parts reusable without changes in different aPS projects, and application-specific parts, which are tailored to a specific aPS and not reusable in a different context without changes.

### B. Software Quality, Software Complexity and Metrics

Generally, there are no universally agreed-upon standards of "good" or "bad" software. Instead, quality is influenced by and depends on different factors and boundary conditions. Still, software quality is specified, measured and evaluated using quality attributes defined in standards, e.g., ISO/IEC 25010. Concerning complexity, it considers aspects like reuse, modularity, changeability and testability. Nevertheless, there is no common understanding of *software complexity* and the various definitions from literature vary greatly. Overall, it is difficult to describe the entire complexity of software directly and comprehensively with a single number [11] and according to [12], it is not directly measurable. Software metrics can be used to compare software units according to specific quality attributes [13], which is achieved by counting code properties and aggregating them according to a fixed algorithm [14]. Regarding software complexity metrics, [15] state that they "do not measure the complexity itself, but instead measure the degree to which those characteristics thought to lead to complexity exist within the code". Thus, although metrics offer consistent, reproducible results, they are not a definite quality judgment but rather indicate possible weaknesses [16].

Many definitions of software complexity relate the term to understanding/comprehending, considering software design and the number of components and their relations [17], or the difficulty/mental effort required for common programming tasks and software comprehension [11], [15], [18–21]. Apart from understanding also maintaining/changing [11], [15], [19–21] and verifying / testing [6], [15], [17], [19–21] software are closely related to software complexity. Based thereon, this paper considers software complexity as the degree to which software is difficult to understand and maintain.

Furthermore, various approaches to categorize software complexity exist, e.g., five classes (cf. Fig. 1) from [15]. The class *Size* is based on the premise that the larger something is, the more difficult it is in terms of understanding. *Control Flow Metrics* assume that program complexity increases with the difficulty of its control flow, e.g., number of forks, while *Information Flow Metrics* target the data exchange between software units via parameters or global variables and rely on the premise that an increasing amount of data shared and a higher number of software modules lead to a higher program complexity. *Software Science Metrics* measure a program's lexical structure and its influence on the program complexity based on the total and the unique number of program tokens, i.e., operands and operators. Finally, *Data Structure Metrics* target the complexity of the data processed by the program.

### C. Software Complexity Measurement Approaches

The use of software metrics is a well-established means in computer science for evaluating software programmed in high-programming languages. One of the simplest and generally accepted complexity metrics is *Lines of Code*, which measures the length or size of a program by counting the lines of a given code file or *Source Lines of Code* (counting non-commentary, non-empty lines only) [22]. McCabe proposed a metric to calculate the *Cyclomatic Complexity*, which is based on the software's control flow graph [23]. In contrast, Halstead suggests counting operands and operators for measuring the length, vocabulary and difficulty of a program [24]. Other works focus on the information flow as a measure for the software complexity [25]. This approach is transferred to control software of lab-sized demonstrators written in high programming languages, e.g., Pascal, in [26], who assume that complexity is composed of a software unit's internal complexity and data exchange with its environment. For measuring the testing effort and understandability, [4] use metrics, which do not consider the software's implemented source code but solely focus on their interfaces and coupling, taking their general functionality into account. Overall, metrics from computer science consider different complexity aspects. However, none of these metrics is applicable to graphical programming languages or considers the boundary conditions, e.g., close connection to hardware, of PLC software. Thus, they are not applicable to IEC 61131-3 software without changes.

Also in the aPS domain metrics for complexity measurement of control software have been developed. Some are based on the introduced metrics from computer science, e.g., [27] transfer the metrics of [23–25] to the IEC 61131-3 language FBD, enlarge it with a size metric to measure the complexity in a safety-critical PLC software and evaluate it with an industrial case study. Another industrially evaluated approach for FBD software, based on [24] and [25], is further considering information flow [16]. Complexity measurement in the language LD was addressed in [28] who target cognitive complexity to measure the understandability and testing complexity, but the approach is not evaluated with industrial software. For assessing the usability and maintainability of PLC software written in SFC, [29] adopt the complexity analysis of flow graphs [23] and develop new metrics, which are evaluated with an academic example. To measure the diagnosability of PLC software written in IL, [5] investigate the applicability of [22–24] to a didactic and an industrial example and highlight the need for future field studies. An adaptation of the metrics from [23] and [24] to all PLC programming languages using a lab-sized demonstrator showed that they are, in principle, suitable for determining the complexity of aPS software [30].

Other authors propose software metrics tailored explicitly to the aPS domain, independent from computer science metrics: [31] transform LD software into petri nets to compare both regarding their design complexity and understandability using a small industrial example. Similarly, [32] propose metrics to measure size, modularity and interconnectedness of LD software and evaluate them with programs of a test-bed.

In summary, software complexity is steadily growing and negatively impacts characteristics like maintenance, testability or reuse. Thus, identifying the most complex software units as a starting point for refactoring would be beneficial to safe costs. But existing PLC software metrics only target one IEC 61131-3 language, apart from the metrics in [30], which have not been applied to industrial control software yet. Further, none of the presented approaches considers all five complexity classes from [15]. Instead, the metrics usually target a specific complexity aspect (cf. Table I). Thus, none of the above-mentioned approaches fulfills all requirements (cf. Sec. II).

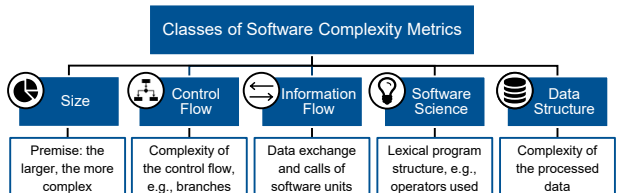
Fig. 1. Overview on software complexity classes from [15].

TABLE I
METRICS FROM LITERATURE WITH TARGET AND CONSIDERED CLASSES

| | Source | Target | | Classes of Software Complexity | | | | |
|---|---|---|---|---|---|---|---|---|
| | | Ind. SW* | IEC Language | Size | Control Flow | Information Flow | Software Science | Data Structure |
| Computer Science | Rosenberg | x | - | x | - | - | - | - |
| | McCabe | x | - | - | x | - | - | - |
| | Halstead | x | - | x | - | - | x | - |
| | Henry, Kafura | x | - | o | - | x | o | - |
| | Phukan et al. | - | - | x | - | x | x | - |
| | Kaur, Singh | - | - | - | - | x | - | x |
| IEC 61131-3 Software | Muslija, Enoui | x | FBD | x | x | x | x | - |
| | Wilch et al. | x | FBD | x | - | x | x | x |
| | Lee, Hsu | o | LD | x | x | - | - | - |
| | Lucas, Tilbury | - | LD | x | - | - | - | - |
| | Kumar et al. | - | LD | x | x | - | x | - |
| | Engell et al. | - | SFC | x | x | - | x | x |
| | Younis, Frey | x | IL | x | x | x | x | - |
| | Capitán et al. | - | all 5 | x | x | x | x | - |
| | This paper | x | 4** | x | x | x | x | x |

Legend: "x" = completely, "o" = partially, "-" = not addressed;
*Ind. SW = Evaluation with Industrial Software examples; ** all but IL

## IV. QUALITATIVE QUESTIONNAIRE STUDY ON COMPLEXITY

To gain insights on complexity measurement, its benefits and challenges from an industrial viewpoint, a small, qualitative questionnaire study was conducted with three market-leading German companies from different industry sectors in the machine and plant manufacturing domain, who apply different strategies for planned reuse of control software but are still challenged by the software's rising complexity. The questionnaire's main aim was to gain insights into the type of complexity, which software developers face during their day-to-day work. The questionnaire consists of seven questions, with five of them being "open questions" with no pre-defined answer options to not bias the software developers with pre-defined options. In all three participating companies, a software developer distributed the questionnaire to his colleagues, gathered the answers and returned them for evaluation. In total, 13 participants answered the questionnaire (3 from company A, 7 from company B, 3 from company C).

For a general assessment, the participants were asked how important they would rate complexity measurement on a scale from 1 (very important) to 4 (not important) and why. Overall, all companies' software developers indicated the measurement of software complexity as important with a mean value of 1.87. Of course, with only 13 participants from three companies, this is not a statistically valid result. However, the participants gave qualitative reasons for their assessment of the importance of complexity measurement and why they consider the use of software metrics as helpful:

- Programmers are "operationally blind" to complexity in their code, as it reflects their own thought processes.
- Early detection of complex parts simplifies refactoring exponentially; later on, often no capacity to fix issues.
- To define testing / the scope of testing.
- Having many programmers with personal, sometimes even conflicting views of complexity requires metrics as an objective method with pre-defined criteria.

Also, the participants were asked to describe the influence of complexity on software characteristics and quality in an open question. Thereby, participants from all companies stated independently the aspects comprehensibility, error-proneness, changeability and reusability (cf. Fig. 2).

Another open question targeted the causes for complexity and the participants mentioned numerous, diverse parameters influencing software complexity (in total about 40 different parameters). Similar answers were summarized, resulting in the following, most mentioned influencing parameters: non-compliance with coding guidelines (stated by 46%, from all companies), extensive functions (46%, mentioned by one company), missing encapsulation/too many functions in one POU (38.5%, all companies), and poor (non-uniform/non-existent) software structure (31%, all companies).

Concerning the five complexity classes described in [15], the participants were asked to rate each class's influence on the overall complexity. Although the impact was estimated quite differently within and between companies, the following tendency emerges: *Information Flow* is rated as most influential, followed by *Control Flow* and *Size*. Additionally mentioned complexity classes, which should be considered, are the choice of programming language, coding guidelines/style guide, commenting and timing characteristics.

Further, the participating software developers provided free-text answers, which illustrated conflicting views on software complexity. For example, one participant mentioned that the different complexity classes and respective metrics should be parameterized in the calculation of the overall complexity to allow an adaptation of the weighing for different stakeholders. Another participant highlighted the balancing act between a fine-granular division of functions into numerous sub-functions and a large function, which is easier to debug. Accordingly, this participant rates complexity as relative, depending on the target, e.g., easy maintenance or easy readability of the software. Another important aspect was the relation of complexity to the number of programmers working within the same software. A participant stated that the more people work on a software project, the easier and more understandable it should be written. Finally, various participants highlighted detecting complex software parts as only the first step, with concepts for subsequently reducing complexity being crucial.

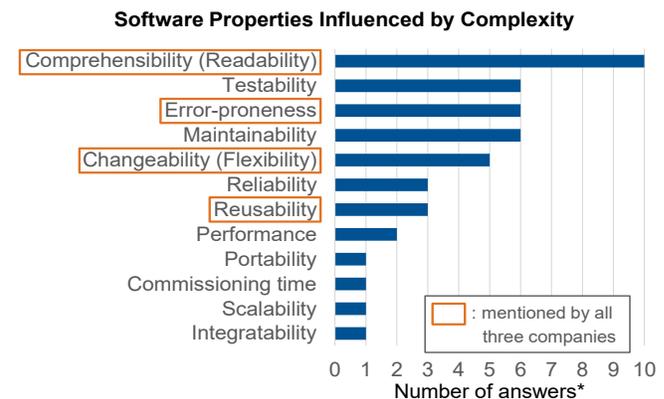

Fig. 2. Qualitative study on the influence of complexity on software characteristics (*open question, similar answers summarized).

In summary, the study qualitatively confirmed that in PLC software development, complexity is a challenging aspect, which might hinder comprehensibility, testability or error-proneness. But between and even within companies differing opinions and estimations on complexity exist, requiring objective means for its assessment. Due to the questionnaire setup with mainly open questions, the developers indicated only aspects, which they thought of at the time of answering. Thus, if a participant did not mention an aspect, it does not imply that it is not important or relevant for his daily tasks. It might merely be the case that the participant did not think of the respective aspect when filling out the questionnaire.

## V. Concept for Complexity Measurement of PLC Software Using Software Metrics

This section introduces the metric concept for complexity measurement including a visualization. As diverse perspectives on complexity exist, different sources were used to develop the concept. These include definitions and metrics from literature, complexity-critical elements of the IEC 61131-3 syntax, feedback from industry experts and an evaluation using industrial software. The concept considers not one individual, single metric, but metrics for all five classes from [15] are included to cover different aspects of complexity and provide comprehensive coverage. Thus, the concept consists of at least one metric for each of the five classes for determining the respective complexity. In total, it contains six metrics. To enable conclusions about the complexity causes, every influencing factor is considered separately. Each considered metric is assigned to exactly one complexity class. Fig. 3 shows a graphical overview of the concept and the calculation of the total complexity divided into three steps (metrics for the different classes, scaling of the values to the corresponding median of each class and, finally, calculating the overall complexity). The used metrics are introduced in the following.

### A. Existing Metrics for IEC 61131-3 from Literature

For four of the introduced classes, five metrics ($M_1$ to $M_5$)

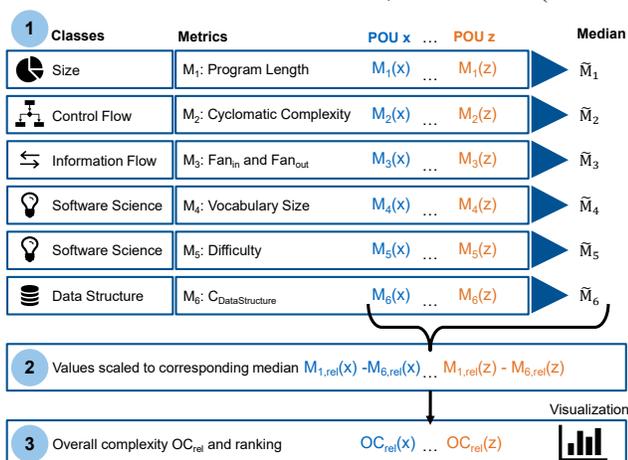

Fig. 3. Concept for aggregating software complexity metrics from different classes to determine the overall complexity.

are taken from literature and adapted to boundary conditions and the syntax of IEC 61131-3, if required. They are briefly introduced below with reference to the class they address.

*Size:* Halstead's Program Length [24] is the sum of all operators ($N_1$) and operands ($N_2$) of a program (1). Operands generally represent data; operators are code elements handling or manipulating operands and producing a result.

$$M_1 := Program\ Length = N_1 + N_2 \quad (1)$$

*Control Flow:* McCabe's Cyclomatic Complexity (*CC*), which can be determined by the number of decision statements (*d*) in a program [23]:

$$M_2 := CC = d + 1 \quad (2)$$

Decision statements for IEC 61131-3 are taken from [30], where (2) was already successfully applied to aPS software.

*Information Flow:* a POU's incoming ($Fan_{in}$) and outgoing ($Fan_{out}$) information flows are multiplied (3). Thereby, incoming information flows are input variables, read access to external data structures and return values of called POUs. Outgoing flows are output variables, writing access to external data structures and passed variables to called POUs.

$$M_3 := FIFO = Fan_{in} * Fan_{out} \quad (3)$$

The metric was used successfully for FBD in [16]. It considers the separation of complexity classes and shows increased stability to the high quantity of inputs/outputs in aPS compared to the original version introduced in [25].

*Software Science:* This class includes the two metrics, Halstead's Vocabulary Size (4) and Difficulty (5) [24].

$$M_4 := Vocabulary\ Size = n_1 + n_2 \quad (4)$$
$$M_5 := Difficulty = \frac{n_1}{2} * \frac{N_2}{n_2} \quad (5)$$

With  $n_1$ ($n_2$): number of unique operators (operands)
  $N_2$: total number of operands
The definition of operators and operands is as in (1).

### B. New Metric Developed for Data Structure Complexity

As there is no suitable metric available to measure the data structure complexity (cf. Table I), a new metric is derived in cooperation with industry experts, who indicate that data structure complexity increases with the number of variables and their type. Thus, three different variable types are distinguished:

- *Interface variables*: Input and output variables of a POU, which are specifically declared.
- *Local variables*: Local variables of a POU such as constants, which are specifically declared.
- *Sub-variables*: All variables included in a multi-element variable, e.g., Structure (if the subordinate variables contain other subordinate variables, these are not considered, as their influence is negligible).

According to industrial feedback, interface variables add more complexity than local variables and sub-variables. Further, elementary data types (e.g., Bool, Int) influence complexity less than multi-element or user-defined variables, which are considered as complex data types. Sub-variables have a low influence regardless of their data type: while they must be considered regarding quantity, structuring variables into multi-element variables should not increase complexity.

Based on these axioms, each variable can be assigned a weighting (cf. Table II). High values indicate a strong influence on complexity, while low values represent a small impact. The data structure complexity ($C_{DataStructure}$) of a POU is calculated from the sum of all its declared variables' weights $w_v$.

$$M_6 := C_{DataStructure}(POU) = \sum_{v \in variables} w_v \quad (6)$$

With *variables*: the set of all declared variables (incl. sub-variables) of the considered POU

### C. Calculation of a POU's Overall Complexity $OC_{rel}$

For each POU, the six complexity metrics ($M_1 - M_6$) are calculated, which leads to six values with different scales per unit (cf. Fig. 3, step 1). To enable analyzing the composition of a POU's complexity and a simple comparison between POUs regarding their overall complexity, the median is used as a reference value. It is chosen as it is stable against outliers and provides reliable results despite the lack of a normal distribution. The median is determined based on the metric values of the POUs to be compared (analyzed sample) for each of the six metrics separately. Then the metric values are set relative to the corresponding median $\widetilde{M}_i$ according to (7), with $i$ ranging from 1 to 6.

$$C_{rel,M_i}(POU) = \frac{M_i(POU)}{\widetilde{M}_i} * 100\% \quad (7)$$

Thereby, $M_i(POU)$ is the calculated value of a specific metric (e.g., $M_1(POU)$ for CC) and the median $\widetilde{M}_i$ is the median of all calculated values for this metric. The result ($C_{rel,M_i}(POU)$) represents the complexity of the respective POU relative to the median of the considered sample. Thus, each POU has six of these values (cf. Fig. 3, step 2). Finally, a POU's overall complexity $OC_{rel}$ is calculated based on the relative complexity values (cf. Fig. 3, step 3) as the sum of the six complexity metrics (cf. (8)). With the weighting factor $w_i$, the increased influence of individual complexity metrics on the overall complexity can be mapped.

$$OC_{rel}(POU) = \sum_{i=1}^{n} w_i * C_{rel,M_i}(POU) \quad (8)$$

With $n$: amount of metrics included (in this case $n = 6$)
Further, the equation (9) has to be fulfilled:

$$\sum_{i=1}^{n} w_i = 1 \quad (9)$$

For most of the languages studied in this paper, the weighting $w_i$ is the same for all metrics ($w_i=1/6$). Only for SFC, a stronger weighting of the metrics CC ($w_i=4/12$) and Halstead's Program Length ($w_i=4/12$) is necessary due to the language-specific characteristics and common use case. In this case, the four other metrics are weighted $w_i=1/12$.

The obtained results are visualized in a bar graph. The bar height indicates the overall complexity $OC_{rel}$ and the POUs are arranged by increasing $OC_{rel}$ along the horizontal axis. The share of the metrics in the overall complexity is shown by the colored subdivisions of the respective bar (cf. Fig. 4). Further, the concept is designed so that any number of metrics can be added or removed. Only the weighting $w_i$ must be adjusted to satisfy (9) if the number of metrics differs.

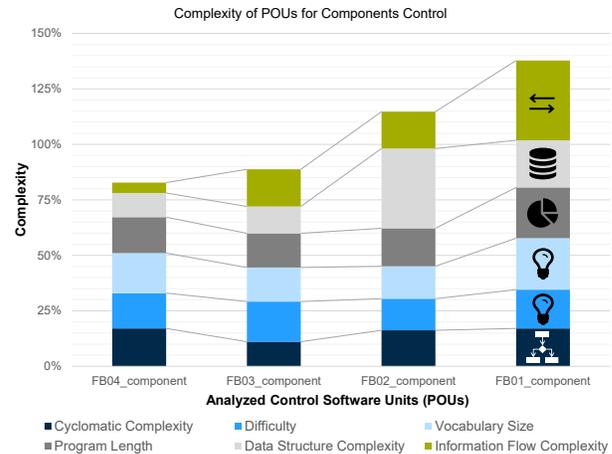

Fig. 4. Overall complexity including results for different classes at the example of POUs implementing component control.

## VI. EVALUATION USING INDUSTRIAL SOFTWARE EXAMPLES IMPLEMENTED IN FOUR IEC 61131-3 LANGUAGES

The complexity measurement concept has been evaluated with industrial control software from two German companies (A and B). The results gained during evaluation and discussions with domain experts are presented. Afterward, findings are discussed in the context of the derived requirements.

### A. Evaluation with Industrial Software Units

The concept evaluation was carried out with 50 POUs from industrial control software projects provided by the two companies A and B. The projects are programmed in the Siemens PLC development environment. The considered POUs implement different functionalities and were programmed using four different languages, including graphical and textual languages (cf. Table III for details).

The software of company A is divided into general parts reusable across different machines, and application-specific parts tailored to a specific machine and customer requirements. Overall, it has a modular structure, which is oriented to the used aPS hardware, e.g., a robot or a turning table. Within these modules (a group of POUs controlling a hardware module), the software is designed and structured in a function-oriented manner with dedicated POUs for specific functionalities. The POUs of company A, which were analyzed in the scope of the evaluation, can be categorized into seven functionality classes (*organizational* for POU calls, *messages* for general communication, *errors* for communication of error messages, *status* of the module, *generic* module functions, *components* for hardware control, and *homing* for moving a module to base position). Apart from application-specific POUs, also selected library POUs, e.g., variants of a POU for controlling pneumatic cylinders, were analyzed.

TABLE III
OVERVIEW ON MANUALLY EVALUATED SOFTWARE UNITS (POUs)

| Programming Language | Company | POU amount and type | Implemented Functionality |
|---|---|---|---|
| Function Block Diagram (FBD) | B | 5 (all FBs) | Safety |
| Sequential Function Chart (SFC) | A | 4 (all FBs) | Homing, Robot control |
| Structured Text (ST) | A | 5 (4 FCs, 1 FB) | Managing order & workpiece data |
| | B | 6 (5 FCs, 1 FB) | Standardized, but adaptable |
| Ladder Diagram (LD) | A | 30 (4 OBs, 26 FBs) | Organizational, messages, errors, status, generic |

TABLE II
WEIGHTS $W_V$ LINKED TO VARIABLE AND DATA TYPE

| Data type | Variable type | | |
|---|---|---|---|
| | Interface variable | Local variable | Sub-variable |
| simple | 3 | 1 | 1 |
| complex | 4 | 2 | 1 |

The plant of company B is controlled by several PLCs, whereby a single PLC is linked to one or multiple machines (of a plant) and the respective process steps. All POUs of company B analyzed in the concept evaluation belong to the same process step. Like company A, the software design is hardware-oriented and structured according to the machine parts into three or four architectural layers. Safety-related parts of the PLC program are implemented in FBD, e.g., safety-related interlocks or fault handling of a component such as a valve. Further, library POUs implementing standardized functionality like mathematical calculations are programmed in ST and can be adapted to application-specific requirements.

The developed complexity metric was applied to all 50 POUs and the results were displayed with the introduced visualization. They were discussed with software developers from companies A and B, who provided feedback regarding the correctness of the calculated values in several expert workshops. The evaluation confirmed the resulting values for the *overall complexity OC$_{rel}$* as a representative metric of the analyzed POUs when comparing their complexity values to each other. Besides, experts agreed that the values of the individual complexity metrics and their distribution represent the complexity classes according to their assessment.

Further, the evaluation showed that POUs implementing the same functionality have similar values regarding the *overall complexity OC$_{rel}$*. They are also similar in the distribution of the individual complexity metrics (cf. Fig. 4, POUs for component control). When ordering all POUs of company A, which are implemented in LD, by increasing complexity values, they were at the same time grouped by functionality classes: in the evaluated sample, POUs with the lowest complexity values are POUs implementing organizational functionalities and POUs with the highest complexity are POUs implementing component control (cf. Fig. 5). Thus, in the sample, POUs for hardware control have the highest complexity, while POUs implementing extra-functional software parts, e.g., messages or errors, show lower complexity values.

Moreover, the analysis of the software units showed that the implemented functionality of a POU is related to its dominant complexity metric: POUs implementing *organizational* tasks show high values for *Software Science Metrics*, while POUs transferring the *status* and, thus, a lot of information for each module, have high complexity values regarding the *Data Structure*. The *Control Flow Metrics* are the dominant metric in POUs implementing *error* handling, while POUs for *component* control have high *Information Flow* values due to the amount of data being transferred from and/or to hardware.

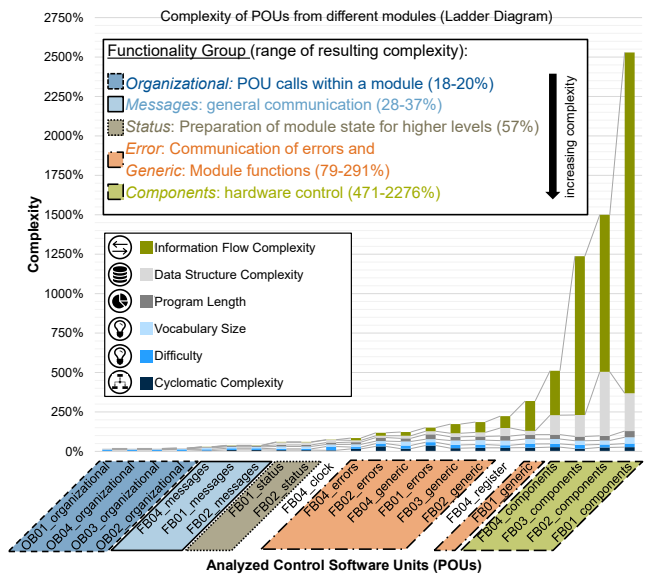

Fig. 5. POUs from different modules sorted by overall complexity and annotated according to their functionality (LD, Company A).

Finally, also POUs providing generic module functionality show a high complexity of the *Information Flow*. Thus, it is assumed that the implemented functionality influences the dominant complexity category.

Regarding the visualization, the interviewed experts from both companies rated the illustration of the calculated metrics results as intuitively understandable and helpful for the interpretation of the determined complexity values. For the analysis, a prototypical implementation was used, which parsed the source files exported from the PLC development environment and enabled the automated metric calculation in Excel. Meanwhile, the prototype has been enlarged to cover the languages LD, FBD, SFC and ST and automatically derive the respective metric values. With the prototype, the *overall complexity OC$_{rel}$* for 2222 POUs contained in six software projects from companies A and B was calculated. Both companies confirmed the obtained complexity values in principle, but a detailed analysis of the results is still pending.

### B. Fulfillment of Requirements and Discussion of Results

The presented concept for software complexity combines different complexity metrics to represent the five introduced complexity classes. These metrics can be weighted and additional metrics can be included if required. Thus, *R1 considering different complexity views* is fulfilled. The concept was developed for and applied to four different IEC 61131-3 languages, including ST (textual) and FBD, LD and SFC (graphical), which fulfills *R2 textual and graphical languages*. The concept was applied successfully to 50 industrial software units implementing different functionalities. However, software units with a great size were only analyzed in LD since the prototypical implementation was available for these at the time of evaluation. Concluding, the *scalability (R3)* is fulfilled in principle but should be examined again in further applications. Regarding the metric calculation effort, the prototype demonstrates that the concept is fully *automatable (R4)*. Finally, the evaluation with industrial experts confirmed that metric results and their visualization are understandable. By splitting the metric value into six sub-values, the developers can derive structures to be revised, although no explicit tasks are provided. Thus, *R5 comprehensibility* is overall fulfilled.

The determined overall complexity $OC_{rel}$ is a representative metric of a POU's complexity in different boundary conditions (evaluated with POUs from two companies differing in their functionality and programming language). However, setting obtained metric values in relation to the sample's median leads to values of more than 100%, which experts rated as confusing, and the concept should be revised in that matter. Although the chosen complexity metrics obtain correct results, the suitability of alternative or additional metrics has to be investigated.

## VII. SUMMARY AND OUTLOOK

This paper presented a literature review on views and definitions of the term software complexity and its calculation using software metrics. A qualitative questionnaire study on the influences and effects of software complexity was conducted. It highlighted the industry's need for support in managing software complexity with objective means. Based on insights from literature, the qualitative study and expert interviews, a concept for assessing software complexity categorized into five classes and applicable to graphical and textual IEC 61131-3 languages was developed. The concept enables identifying the most complex POUs within an analyzed sample, including their complexity composition, which allows determining the main cause for the POU's high complexity rating and a suitable starting point for refactoring. It was successfully validated with 50 POUs from two companies and automated in a prototype.

In future work, the metric values will be optimized to avoid values of more than 100%. For evaluation purposes, the concept should be validated with additional POUs from different companies and be enlarged with other classes, e.g., time behavior or comments. Insights gained from the qualitative study and expert interviews can be used to design and conduct a larger questionnaire study to obtain statistically meaningful results regarding challenges of control software complexity. The proposed metric $OC_{rel}$ enables comparing POUs before and after a refactoring process as a first step for estimating the costs caused by complexity during software development. Further, different stakeholders, e.g., commissioners and maintenance staff, might have different perceptions regarding complexity and require other classes, which will be investigated.

## REFERENCES

[1] B. Vogel-Heuser, A. Fay, I. Schaefer, and M. Tichy, "Evolution of software in automated production systems: Challenges and research directions," *J. Syst. Softw.*, vol. 110, pp. 54–84, 2015.
[2] B. Vogel-Heuser, J. Fischer, S. Rösch, S. Feldmann, and S. Ulewicz, "Challenges for maintenance of PLC-software and its related hardware for automated production systems: Selected industrial Case Studies," in *IEEE ICSME*, Bremen, Germany, 2015, pp. 362–371.
[3] K. Thramboulidis and A. Buda, "3+1 SysML view model for IEC61499 Function Block control systems," in *IEEE INDIN*, Osaka, Japan, 2010, pp. 175–180.
[4] N. Kaur and A. Singh, "A Complexity Metric for Black Box Components," *Int. J. Soft Comp. Eng.*, vol. 3, no. 2, 179-184, 2013.
[5] M. B. Younis and G. Frey, "Software quality measures to determine the diagnosability of PLC applications," in *IEEE ETFA*, Patras, Greece, 2007, pp. 368–375.
[6] C. V. Ramamoorthy, W. T. Tsai, T. Yamaura, and A. Bhide, "Metrics Guided Methodology," in *IEEE COMPSAC*, Chicago, 1985, pp. 111–120.
[7] H. Prähofer, F. Angerer, R. Ramler, and F. Grillenberger, "Static Code Analysis of IEC 61131-3 Programs: Comprehensive Tool Support and Experiences from Large-Scale Industrial Application," *IEEE Trans. Ind. Informat.*, vol. 13, no. 1, pp. 37–47, 2017.
[8] B. Vogel-Heuser, J. Fischer, and E.-M. Neumann, "Goal-Lever-Indicator-Principle to Derive Recommendations for Improving IEC 61131-3 Control Software," in *IEEE IEEM*, Singapore, 2020, pp. 1131–1136.
[9] Siemens AG, *Standards compliance according to IEC 61131-3 (3rd Edition): Function Manual.* [Online]. Available: https://cache.industry.siemens.com/dl/files/748/109476748/att_845621/v1/IEC_61131_compliance_en_US.pdf (accessed: Jan. 17 2021).
[10] B. Meyer, *Object-oriented software construction,* 2nd ed. Upper Saddle River, NJ: Prentice Hall PTR, 2009.
[11] H. Zuse, *Software complexity: Measures and methods*. Berlin: De Gruyter, 1991.
[12] M. Shepperd, "An evaluation of software product metrics," *Information and Software Technology*, vol. 30, no. 3, pp. 177–188, 1988.
[13] L. Sonnleithner and A. Zoitl, "A Software Measure for IEC 61499 Basic Function Blocks," in *IEEE ETFA*, Vienna, Austria, 2020, pp. 997–1000.
[14] E. J. Weyuker, "Evaluating software complexity measures," *IEEE Trans. Softw. Eng.*, vol. 14, no. 9, pp. 1357–1365, 1988.
[15] A. Lake and C. R. Cook, "Use of Factor Analysis to Develop OOP Software Complexity Metrics," 1994.
[16] J. Wilch *et al.,* "Introduction and Evaluation of Complexity Metrics for Network-based, Graphical IEC 61131-3 Programming Languages," in *IEEE IECON*, Lisabon, Portugal, 2019, pp. 417–423.
[17] *Systems and software engineering -- Vocabulary*, ISO/IEC/IEEE 24765:2010(E), Institute of Electrical and Electronics Engineers, Piscataway, NJ, USA, 2010.
[18] J. E. Sullivan, "Measuring the Complexity of Computer Software," NTIS AD-A007 770, 1975.
[19] V. R. Basili, "Tutorial on Models and Metrics for Software Management and Engineering," in *IEEE Comp. Softw. Appl. Conf.*, Chicago, 1980.
[20] W. Harrison, K. Magel, R. Kluczny, and A. DeKock, "Applying software complexity metrics to program maintenance," *Computer*, vol. 15, no. 9, pp. 65–79, 1982.
[21] B. Curtis, "In search of software complexity," in *W. on Quantit. Softw. Models for reliability, complexity & cost*, New York, 1979, pp. 95–106.
[22] J. Rosenberg, "Some misconceptions about lines of code," in *Int. Softw. Metrics Symp.*, Albuquerque, NM, USA, 1997, pp. 137–142.
[23] T. J. McCabe, "A Complexity Measure," *IEEE Trans. Softw. Eng.*, SE-2, no. 4, pp. 308–320, 1976.
[24] M. H. Halstead, *Elements of software science*. New York, NY, USA: Elsevier Science Inc., 1977.
[25] S. Henry and D. Kafura, "Software Structure Metrics Based on Information Flow," *IEEE Trans. Softw. Eng.*, SE-7, no. 5, pp. 510–518, 1981.
[26] A. Phukan, M. Kalava, and V. Prabhu, "Complexity metrics for manufacturing control architectures based on software and information flow," *Comput. & Industr. Eng.*, vol. 49, no. 1, pp. 1–20, 2005.
[27] A. Muslija and E. P. Enoiu, "On the Measurement of Software Complexity for PLC Industrial Control Systems using TIQVA," in *ACM/SIGAPP SAC*, Brno Czech Republic, 2020, pp. 1556–1565.
[28] L. Kumar, R. Jetley, and A. Sureka, "Source code metrics for programmable logic controller (PLC) ladder diagram (LD) visual programming language," in *IEEE WETSoM*, Texas, 2016, pp. 15–21.
[29] S. Engell, A. Dandachi, and S. Lohmann, "Impact of Complexity on Logic Controller Design," *IFAC Proc. Vol.*, vol. 40, no. 6, pp. 121–126, 2007.
[30] L. Capitan and B. Vogel-Heuser, "Metrics for software quality in automated production systems as an indicator for technical debt," in *IEEE CASE*, Xi'an, 2017, pp. 709–716.
[31] J.-S. Lee and P.-L. Hsu, "A new approach to evaluate ladder logic diagrams and Petri nets via the IF-THEN transformation," in *IEEE SMC. e-Syst. & e-Man for Cybern. in Cybersp.*, Tucson, 2001, pp. 2711–2716.
[32] M. R. Lucas and D. M. Tilbury, "Methods of measuring the size and complexity of PLC programs in different logic control design methodologies," *J. Adv. Manuf. Techn.*, vol. 26, 5-6, pp. 436–447, 2005.